\documentstyle[11pt,newpasp,twoside,epsf]{article}
\def\edcomment#1{\iffalse\marginpar{\raggedright\sl#1\/}\else\relax\fi}
\reversemarginpar

\begin{document}
\title{Comparing Chemical Abundances of the 
Damped Ly$\alpha$\ Systems and Metal-Poor Stars} 
 \author{Jason X. Prochaska}
\affil{UCO/Lick Observatory, UC Santa Cruz, Santa Cruz, CA, 95064}

\begin{abstract}
I briefly draw comparisons between the fields of damped Ly$\alpha$
and metal-poor stellar abundances.  In particular, I examine their 
complementary age-metallicity relations and comparisons between 
the damped Ly$\alpha$ and dwarf galaxy abundance patterns.  
Regarding the latter, I describe a series of problems concerning
associating high $z$ damped Ly$\alpha$ systems with present-day dwarfs.
\end{abstract}

\section{Introduction}

I wish to first acknowledge the wisdom of the organizers for bringing the
damped Ly$\alpha$ and metal-poor stellar abundance communities together in a 
Joint Discussion aimed at heightening 
communication and collaboration between the two fields.  
While previous conferences on chemical abundances have included members
of each group, talks were generally organized such that we have talked
to one another instead of with one another.
From the perspective of a DLA researcher, the proceeding 50~years of
stellar abundance research is invaluable to drawing interpretations
on nucleosynthesis from DLA observations.  I suspect that as DLA abundance
studies become a mature field, the stellar community will similarly gain
from observations of these young, metal-poor galaxies. 

The organizers charged
me with reviewing the fields to open the JD.  In Sydney, I briefly
compared the observations, techniques of analysis, and major systematic
uncertainties in each field.  I then described a few areas of research where
the fields clearly intersect and where a joint analysis impacts
theories of chemical evolution, nucleosynthesis, and galaxy formation.
In this proceeding I present an even more brief summary. 

\section{Observations and Analysis}

Presently, the data for damped Ly$\alpha$ and metal-poor stellar abundance
research is acquired with the same telescopes on the same instruments
and with comparable spectral resolution.   Ionic transitions are analysed
in a similar fashion:  integrated 
equivalent widths or column densities are derived from isolated, 'clean' lines
while spectra synthesis or line-profile fitting techniques are applied
to blended and crowded regions.  The derivation of elemental abundances
from the ionic measurements, however, follows different paths.  For
stellar abundances, one introduces a stellar atmosphere derived from the
spectroscopic and/or photometric observations of the star.  The atmosphere
is typically parameterized by three parameters -- temperature, gravity, and 
microturbulence -- and one solves the radiative transfer equations to 
predict equivalent widths and/or spectral profiles.  The uncertainties 
attributed to this modeling (e.g.\ non-LTE corrections; see Apslund's contribution) 
generally dominate the statistical errors associated with the observations.

With the DLA systems, one must consider ionization corrections to convert
the measured ionic column densities into elemental
abundances.  In practice, the corrections are generally assumed to be small.
This assumption is supported by theory and observation, although with exception
(see Vladilo et al.\ 2001; Prochaska et al.\ 2002).
The most important systematic error in the DLA analysis is dust depletion;
the DLA observations provide gas-phase abundances which may significantly
underestimate the true abundances of refractory elements like Fe, Cr, and Ni
which deplete from the gas-phase onto dust grains (e.g.\ Jenkins 1987).
This effect dominates the uncertainty of DLA abundance studies and
plays a central role in nearly all interpretation drawn from the observations.

\section{Age-Metallicity Relation (AMR)}

One research area where the two fields nicely complement each other
is in describing an age-metallicity relation (AMR).  
%In stellar research,
%researchers have focused primarily on Galactic disk stars to determine its
%chemical evolution history (e.g.\ Edvaardsson et al.\ 1993; 
%Rocha-Pinto et al. 2000).   Because of the systematic uncertainties relating
%to isochrone fitting,  stellar ages have absolute uncertainties of at least
%1~Gyr and, in particular, it is difficult to distinguish between ages of 
%10-15~Gyr.  Therefore, the AMR described from stellar observations best
%describes metallicity evolution over the last 10~Gyr of the universe.
%In contarst, the ages of the DLA are precisely determined by combining
%their cosmological redshift with an assumed cosmology.  If we adopt the 
%concordance cosmology (e.g.\ Spergel et al.\ 2003), the age is accurate
%to a few 100~Myr and relative ages have significnatly higher precision.
Stellar research has focused primarily on Galactic disk stars and therefore
probes the chemical enrichment history of a single galaxy over the past
$\approx 10$~Gyr.  Because of the uncertainties of isochrone fitting, it is
difficult to determine absolute ages with meaningful precision for ages
$> 10$~Gyr.  In contrast, the DLA observations reveal the AMR
of the population of high $z$ galaxies which dominate the H\,I content of the
universe.  Therefore, the DLA trace the `cosmic' mean metallicity in neutral
gas (e.g.\ Prochaska et al.\ 2003).  
The ages of the DLA systems are precisely calculated by combining
their cosmological redshift with the `concordance cosmology' 
(e.g.\ Spergel et al.\ 2003).  
The challenge with DLA studies is pursuing the
AMR to $z<1.7$ because space-bourne UV observations
are required to observe the Ly$\alpha$ profile. 
Therefore, the DLA and stellar abundance measurements
complement one another both temporally (spanning nearly the entire history
of the universe) and spatially (examining a single galaxy to a population of
galaxies).

\begin{figure}[t]
%\plotfiddle{prochaska.fig1.ps}{2.4in}{270}{35}{35}{-135}{200}
\plotfiddle{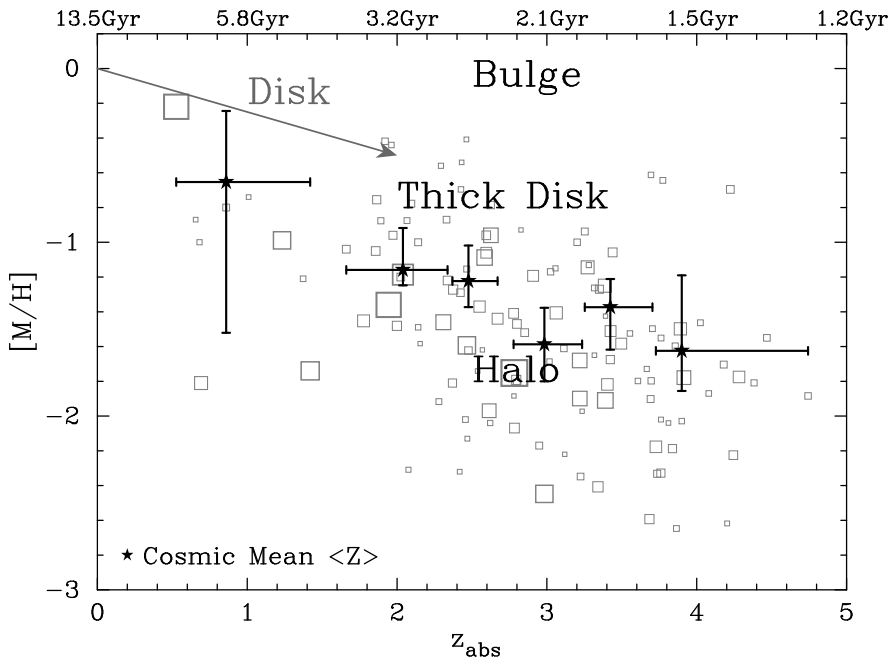}{3.1in}{270}{45}{45}{-180}{255}
%\plotone{fig1.ps}
\caption{Age-metallicity relations for the DLA systems (summarized by
the individual and binned data points) and the Galactic disk (solid arrow).  The
rough locations for the Galactic thick disk, halo, and bulge are also shown in
the figure.  
}
\end{figure}

For very different reasons, the AMR's measured from Galactic stars and the
DLA systems have had controversial histories.  Indeed, my own studies have
contributed to the controversy surrounding the AMR derived for the DLA systems
(see also Kulkarni, this proceedings).
In this case, the analysis has been limited by uncertainty relating to small 
sample size;  even with 50~DLA systems 
there was no statistically significant evolution in the
cosmic metallicity.  Similarly, the AMR of the
Galactic disk remains a point of great debate with competing groups
arguing for and against any trend of metallicity with age (Edvardsson et al.\ 1993;
Rocha-Pinto et al.\ 2000; Feltzing, Holmberg, \& Hurley 2001;
Ibukiyama \& Arimoto 2002).
It remains unclear to me whether the debate revolves around sample bias
or age determinations.

These uncertainties aside, 
it is illustrative to compare AMR's taken from the two fields of research.
In Figure~1, I present a census of DLA age-metallicity observations
overplotted with one estimate of the Galactic AMR 
(Rocha-Pinto et al.\ 2000) and rough estimates
for the age and metallicity of the Galactic bulge, thick disk, and halo
stellar components.  
Clearly, the DLA systems at $z>2$ are distinct in both metallicity and
age from the Galactic disk AMR.  As first noted by Pettini et al.\ (1997), 
the majority of DLA metallicities lie between the peaks of the
distributions for the thick disk and halo populations.
This indicates the majority of gas associated with the DLA systems at $z>2$
may feed the formation of the thick disk and halo populations but has
insufficient metallicity at these epochs
to produce the majority of disk stars.

As the AMR's are refined and extended to include additional stellar
populations and local galaxies as well DLA systems at $z<1$, one
will gain further insight into the nature of the DLA systems and
their connection to modern galaxies.  Future surveys will reveal
whether a sample of metal-rich DLA exist at $z \sim 1.5$ which
can be identified as the gas reservoirs of spiral disk formation.
The identification of this gas is an important aspect of tracing the
history of disk formation.
In addition, age-metallicity measurements for high $z$ galaxies
(e.g.\ LBG, ERO's) will allow comparisons with the DLA and stellar populations,
illuminating the processes of galaxy enrichment during the early universe.

\begin{figure}[t]
%\plotfiddle{prochaska.fig2.ps}{2.4in}{270}{38}{35}{-145}{200}
\plotfiddle{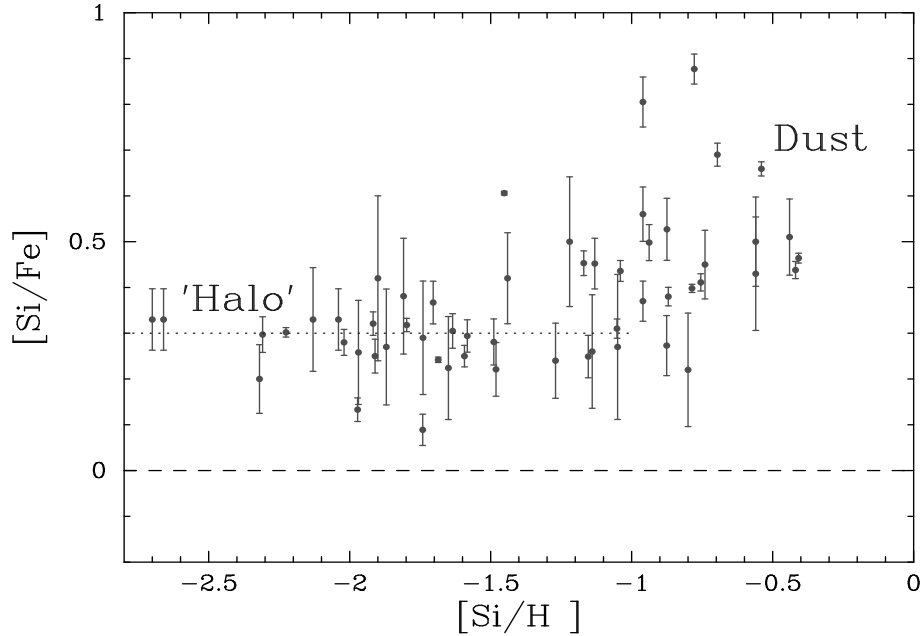}{3.1in}{270}{48}{45}{-190}{255}
\caption{Gas-phase Si/Fe measurements for 56 DLA systems observed with
echelle spectroscopy.   At low metallicity, the Si/Fe values exhibit a
`plateau' at [Si/Fe]~$\approx +0.3$~dex which matches the nominal value
for the Galactic halo (dashed line; McWilliam et al.\ 1995).  
At higher metallicity,
however, the Si/Fe values increase because of the effects of differential
depletion (Fe is more readily depleted from the gas-phase onto dust grains).
}
\label{fig:SiFe}
\end{figure}

\section{DLA systems, dwarf galaxies, and the Galactic halo}

One of the most exciting aspects of connecting the fields of DLA and
stellar abundance research is to draw comparisons between abundance
patterns of local stellar populations and those observed in the early
universe.  Indeed, this was the focus of several talks and posters of this JD
(e.g.\ Tolstoy, Bonifacacio, Dessauges-Zavadsky).  A
principal challenge to drawing such comparisons is uncertainty in the
DLA abundance patterns related to differential depletion. 
This point is well illustrated by Figure~2 where I present
gas-phase Si/Fe ratios against metallicity for a sample of 56 DLA systems
with echelle observations.
As noted by Prochaska \& Wolfe (2002), the metal-poor DLA systems
exhibit [Si/Fe]
values comparable to the Galactic halo with remarkably small scatter
(particularly given that each data point 
represents a unique galaxy with its own specific
chemical enrichment history).  This Si/Fe plateau suggests the metal-poor
DLA systems have been predominantly enriched by Type~II supernovae, perhaps in the
same manner as the Galactic halo.  At higher metallicity, however,
the Si/Fe ratios {\it increase}, in contradiction with any empirical or
theoretical trends of chemical evolution.  The rise in Si/Fe is explained
by differential depletion; higher metallicity DLA systems have significant
dust-to-gas ratios and a substantial fraction of their Fe is 
locked into dust grains.
If one explains the rising Si/Fe ratios in terms of differential depletion,
it raises the possibility that the enhancement of Si/Fe
at all metallicity is dominated by differential depletion. 
Presumably, this interpretation would 
require an uncomfortable degree of fine-tuning to reproduce a [Si/Fe] plateau
at [Si/H]~$<-1$ which matches the nucleosynthetic enhancement of Si/Fe
observed for the Galactic halo, yet more peculiar coincidences exist.

Presently, the majority of both communities favor imposing 
significant depletion corrections to the observed DLA abundances
(e.g.\ Vladilo 2002).   To estimate the corrections, 
one must assume the intrinsic value for the relative abundance of a refractory
and non-refractory element and also assume a differential depletion pattern.
Standard practice is to adopt an empirical depletion pattern based on ISM
observations (e.g.\ Savage \& Sembach 1996) and to
assume [Zn/Fe]~$\approx 0$ intrinsically (i.e.\ attribute
any departures from solar as depletion).
Under this assumption, 
one finds that the majority of the DLA systems show nearly solar
relative abundances at nearly all metallicity and redshift where Zn is observed.
This leads to the conclusion that the relative abundances of the metal-poor DLA
do not match Galactic metal-poor stars, specifically the enhanced $\alpha$/Fe
ratios observed for these stars (note the contradiction with my 
interpretation of Figure~2).  
Instead, these researches argue the DLA abundance patterns more 
closely resemble recent results for 
dSph and dIrr galaxies (e.g.\ Shetrone et al.\ 2001; 
Venn et al.\ 2003; Tolstoy et al.\ 2003).

At first glance, the correspondence between the DLA systems and dwarf
galaxies is a comforting picture.  In hierarchical cosmology, galaxies
at $z \sim 2$ are actively merging to build-up our present-day galaxies.
The DLA systems are identified with `protogalactic clumps'
within dark matter halos (Haehnelt et al.\ 1998; Maller et al.\ 2001)
whose masses might be comparable to modern dSph and dIrr galaxies.
For the following reasons, however, I contend that connecting the DLA
systems to modern dwarf galaxies is premature and potentially in great
conflict with hierarchical cosmology. Consider:

%\begin{itemize}

\noindent
$\bullet$ In hierarchical cosmology, the majority of DLA systems 
will not evolve into present-day dwarf galaxies.

%As noted above, in hierarchical cosmology the DLA systems correspond to 

In order to explain the kinematic characteristics 
observed for the DLA systems (Prochaska \& Wolfe 1997) within
the paradigm of hierarchical cosmology, the `protogalactic clumps' identified
as DLA systems must arise predominantly in galaxies with $v_c > 125$~km/s
(e.g.\ Maller et al.\ 2001).  Furthermore, these clumps are actively
merging with one another to build up the central galaxy of the dark
matter halo.  In this scenario, only a small fraction of the DLA population
(presumably the low mass tail) could serve as the progenitors of present-day
dwarfs;  the majority will be merged into larger galaxies by $z \sim 1$.
Therefore, any correspondence between the abundance patterns of the 
$z \sim 2$ DLA systems and present-day dwarfs may have to be considered
a coincidence.

\noindent
$\bullet$ Where is the gas enriched by Type~II SN?  

An important challenge raised by observers studying the elemental
abundances of dSph in the Local Group is that these stars 
exhibit abundances which are very different from those measured for
the Galactic halo (specifically, the Galactic halo within a few
kpc of the Sun).  This apparently contradicts the favored scenario 
for the formation of the Galactic halo in hierarchical cosmology, i.e., 
it formed from the accretion of dwarf satellites during
the first few Gyr of the universe.  One possible (and fashionable)
resolution of this problem is that the dwarf galaxies which exist
today (i.e.\ those which have not yet merged with the Milky Way) have
had different chemical enrichment histories from those which
merged to form the Galactic halo.  As noted above, however, in CDM cosmology
it is the DLA systems which correspond to the merging dwarfs.  
This identification leads to a more significant conflict regarding
the origin of Galactic metal-poor stars.
If we interpret the DLA abundance patterns as matching present-day 
dwarf galaxies, one is left with the problem: ``Where is the gas enriched by
Type~II SN in the early universe which fueled the formation of the
metal-poor stars of the Milky Way?''

%\item The DLA galaxies are gas-rich whereas the dSph galaxies are gas-poor.

%Although it is unkwon when or even how dwarf spheroidal galaxies lost
%their gas, it should be stressed that they have insufficient H\,I surface
%density to reprsent a DLA system.  There are indications that dSph galaxies
%underwent multiple `burst' episodes of star formation in the pas and
%presumably they had large H\,I surface densities during or just prior
%to these episodes.  To link present-day dSph galaxies with the DLA systems,
%one requires that hte duty cycle of the star forming episode is long 
%($>$~few Myr) which may contradict the starburst interpretation.

\noindent
$\bullet$ Ages, SFR, Zn, etc.

There are a number of other issues which may contradict or at least
complicate the dwarf/DLA connection: (1) the ages of the DLA galaxies
are too young to be consistent with the slow, steady SFH generally associated
with dIrr galaxies; (2) SFR's derived for the DLA systems from the
C\,II$^*$ absorption are typical of those expected for spiral disk 
galaxies (Wolfe, this proceedings); 
(3) the DLA galaxies are gas-rich whereas the dSph galaxies are gas-poor.
To link the two populations, one may require that the duty cycle for
star formation in the dSph galaxies was significantly longer than the
starburst behavior suggested by their stellar populations; and
(4) the reported agreement between DLA and dwarf abundance patterns
hinges on the assumption
that [Zn/Fe]=0 intrinsically.  If [Zn/Fe] is even $+0.2$~dex, 
then the DLA observations would be in good agreement with the 
Galactic halo abundance patterns.  I note that several new issues
related to the [Zn/Fe]=0 presumption were raised at the JD
by Nissen, Apslund, and Israelian. 

%\end{itemize}

These issues aside, comparisons of the DLA and metal-poor stellar abundance
patterns offer a powerful and insightful means of studying nucleosynthesis
and galaxy formation.  I am confident that ongoing
studies of stellar populations
in the Milky Way and Local Group as well as more comprehensive analysis
of the DLA systems (e.g.\ Prochaska, Howk, \& Wolfe 2003; Dessauges-Zavadsky,
this proceeding) will lead to new puzzles and discoveries over the
next years.

\begin{references}

\reference
Edvardsson, B., Andersen, J., Gustafsson, B. et al. 1993, \aap, 275, 101 

\reference
Feltzing, S., Holmberg, J., \& Hurley, J.R. 2001, \aap, 377, 911

\reference
Haehnelt, M.G., Steinmetz, M. \& Rauch, M. 1998, \apj, 495, 647

\reference
Ibukiyama, A. \& Arimoto, N. 2002, \aap, 394, 927

\reference
Jenkins, E.B. 1987, in Interstellar Processes, ed.
D.J. Hollenbach and H.A. Thronson, Jr. (Boston: D. Reidel Publishing
Company), p. 533

\reference
Maller, A.H., Prochaska, J.X., Somerville, R.S., \& Primack, J.R.
2001, \mnras, 326, 1475

\reference
McWilliam, A., Preston, G.W., Sneden, C., \& Searle,L. 1995, \aj,
109, 2757

\reference
Pettini, M., Smith, L.J., King, D.L., \& Hunstead, R.W. 1997, \apj, 486, 665 

\reference
Prochaska, J.X., Howk, J.C., O'Meara, J.M. et al. 2002, \apj, 571, 693
%, Tytler, D., Wolfe, A.M., Kirkman, D., Lubin, D., \& Suzuki, N. 

\reference
Prochaska, J.X., Gawiser, E., Wolfe  et al. 2003, \apjl, 595, L9
%A.M., Castro, S., \& Djorgovski, S.G. 

\reference
Prochaska, J.X., Howk, J.C., \& Wolfe, A.M. 2003, Nature, 423, 57

\reference
Prochaska, J.X. \& Wolfe, A. M. 1997, \apj, 486, 73

\reference
Prochaska, J.X. \& Wolfe, A.M. 2002, \apj, 566, 68 

\reference
Rocha-Pinto, H.J., Scalo, J., Maciel, W.J., \& Flynn, C. 2000,
\aap, 358, 869

\reference
Savage, B. D. and Sembach, K. R. 1996, ARA\&A, 34, 279

\reference
Shetrone, M., C\^ot\'e, P., Sargent, W.L.W. 2001, \apj, 548, 592 

\reference
Spergel, D.,N., et al.\ 2003, \apjs, 148, 175

\reference
Tolstoy, E., Venn, K.A., Shetrone, M. et al. 2003, \aj, 125, 707
%, Primas, F., Hill, V., Kaufer, A., \& Szeifert, T. 

Venn, K.A., Tolstoy, E., Kaufer, A. et al. 2003, \aj, 126, 1326
%, Skillman, E.D., Clarkson, S.M., Smartt, S.J., Lennon, D.J., \& Kudritzki, R.P.

\reference
Vladilo, G., Centuri\'on, M., Bonifacio, P., \& Howk, J.C. 2001, \apj, 557, 1007

\reference
Vladilo, G. 2002, \aap, 391, 407

%Wolfe, A. M., Prochaska, J.X., \& Gawiser, E. 2003, \apj, 593, 215\\

\end {references}

\end{document}